# Interval Elicitation of Forecasts in a Prediction Market Reveals Lack of Anchoring "Bias"

KENNETH C. OLSON, CHARLES R. TWARDY, AND KATHRYN LASKEY, George Mason University
MARK A. BURGMAN, University of Melbourne

## 1. INTRODUCTION

DAGGRE was part of a program that began in 2010 as a multi-team forecasting challenge sponsored by IARPA. To harness the "wisdom of crowds", researchers built an online combinatorial prediction market for predicting events of interest to the U.S. intelligence community. Prediction markets have a long history (See Rhode & Stumpf, 2004.), and their value is described at length by Surowiecki (2004). The DAGGRE market possessed an interface to allow users to input probability judgments directly.

In three-point elicitation (TPE) of judgment, people provide their highest realistic estimates, lowest realistic estimates, and best estimates of a quantity. The structured question format results in more accurate judgments than other formats (Speirs-Bridge et al., 2010). TPE of forecasts can have advantages over a market with single point elicitation, especially improvement of calibration. Soll and Klayman (2004) showed that intervals constructed by TPE were more likely to contain the true value in question. However, TPE can also have disadvantages, such as requiring more effort from participants. The requirement of two extra responses might discourage forecasts. However, any advantage one method has over the other might be shared by a hybrid method. Hence, the experiment described herein required that some forecasters use TPE whenever they wished to edit a market estimate for the first time on the DAGGRE prediction market.

We expected that requiring TPE before viewing and editing of the market for an event would have the added advantage of increasing the diversity of opinion in the market, a major benefit according to the wisdom-of-the-crowd view (Surowiecki, 2004). In support of this view, Scott Page has developed computer models to demonstrate the benefits of cognitive diversity in group judgment (2007).

We had two reasons to believe that TPE of participants prior to accessing the market history would improve edits to the market. The first reason, as Klayman, Soll, Juslin, et al. (2006) suggest, is that more samples of a person's knowledge provide more opportunities to capture the truth. Furthermore, by answering several questions, a judge can assess several different hypotheses. This fits nicely with literature on a psychological debiasing technique called "consider-the-opposite" (Koriat, Lichtenstein, & Fischhoff, 1980). Several studies (e.g., Einhorn & Hogarth, 1978) point to the need for people to consider evidence for what initially seem like implausible outcomes.

The second reason to believe that TPE would improve edits is that probability elicitation up front should reduce mental anchoring on any previous market estimates. Anchoring is a bias first described by Tversky and Kahneman (1974) that can cause irrelevant information to influence a person's judgment. It is assumed that people start with implicit or explicit anchors and make incremental adjustments to reach their estimates. The adjustments are usually insufficient so that the anchor has undue influence on final estimates. Anchoring with insufficient adjustment has proven extremely difficult to correct (e.g., Wilson, Houston, Etling, et al., 1996).

We suspected that participants in the DAGGRE market who could immediately see the estimate history for the probability of an event would anchor on what they saw in that history and fail to adjust away from it when making their own edits. The most likely anchor seemed to be the current state of the market. That this is not irrelevant information is important to note. Anchoring might not constitute a typical cognitive bias in this situation because the anchor is potentially informative.

While TPE alone might encourage more adjustments to estimates, the difficulty of avoiding anchoring bias indicates that a better strategy would be to remove the suspected anchor from view. Therefore, the experiment was designed to elicit probability estimates before participants could access the market history. We hypothesized that self-generated estimates would serve as reference points rather than the current market estimate.

Combining TPE with the prediction market could generate more accurate aggregate forecasts than either alone. We hypothesized that people asked to do TPE would show more accurate judgment when they edited the market. Although their initial estimates could be extremely inaccurate, they could later integrate their own judgment with the market history to make edits to the market that would improve its overall accuracy. However, because TPE is more laborious for participants, we further hypothesized that it would reduce their level of participation in the market.

## 2. EXPERIMENT

Half of forecasters were randomly assigned to a new TPE interface, while the other half were assigned to a regular single point interface. In the TPE condition, people provided highest, lowest, and best estimates of the probability of the event in question. After they clicked on "Commit", they could see the distribution of responses in the TPE condition and the history of market edits for the question. Only after entering their three-point estimate were they able to edit the market themselves.

## 3. RESULTS

Of the 1275 registrants for the TPE condition, 412 made at least one estimate on a question that had resolved by the close of the prediction market. 80 of the 412 made at least one edit to a question. Of the 1276 registrants for the regular condition, 145 made at least one edit to a question. The difference between the number of active participants in the TPE and regular conditions and the difference between the number of participants in the TPE condition who made estimates and those who made both estimates and edits were statistically significant. We conclude that the TPE condition typically increased a person's initial participation, but ultimately decreased participation.

There was a significant difference between the best estimates and first edits on a question from people in the TPE condition, which suggests that they were not anchoring on the state of the market when providing initial estimates but were moving closer to the state of the market when making edits. Furthermore, there was not a significant difference between the first edits of the TPE condition and the first edits of the regular condition, which suggests that people in the regular condition were responding similarly to people in the TPE condition when making edits.

|                                | Mean Brier Score (95% CI) |
| ------------------------------ | ------------------------- |
| Best of 3-Pt                   | 0.74 (0.64, 0.84)         |
| Closest Regular group edit     | 0.69 (0.55, 0.83)         |
| First Edit 3-Pt group market   | 0.60 (0.45, 0.75)         |
| First Edit Regular group market | 0.71 (0.60, 0.82)        |

Table 1. Accuracy per active participant. Brier scores are similar for different types of forecasts.

Accuracy of responses can be assessed in two main ways: average accuracy per person in a condition and accuracy of an aggregated estimate over people's responses in a condition. For the former, Table 1 shows the means and confidence intervals of Brier scores. The similarity of results in the first two rows negates the possibility that people in the TPE condition were timing their estimates for when available information could improve their accuracy relative to estimates in the regular condition. The edits in the regular condition that were closest in time to the estimates in the three-point condition were slightly more accurate. Furthermore, as presented in the last two rows, edits from people in the TPE condition were more accurate that edits from people in the regular condition. Although the pattern of results support our hypotheses, no differences were statistically significant.

For assessing the accuracy of an aggregated estimate over people's responses, several methods were tried. Ultimately, the most recent estimates or edits prior to 8:30 AM each day were recorded and scored; the Brier scores were averaged over days and then questions. We were most interested in whether the edits of participants in the TPE condition would be better than the edits of participants in the regular condition (dashed blue line and solid red line, respectively, in Figure 1). Although they were all in the same prediction market, we treated the aggregated estimates in isolation. There were no significant differences between the conditions.

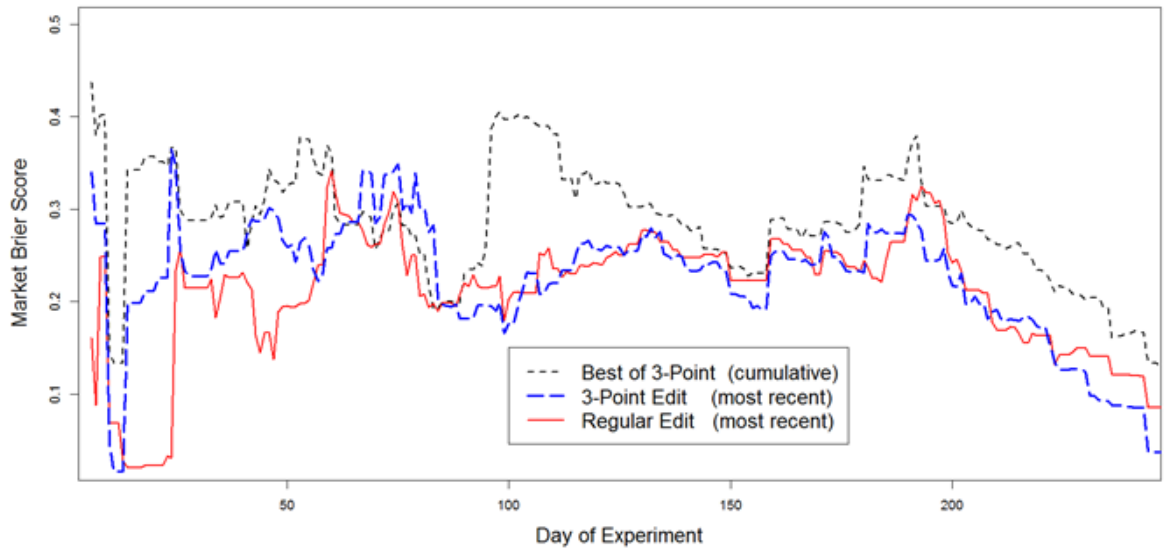

Fig. 1. Average daily accuracy for three aggregation methods. Brier scores were averaged over forecasting problems that resolved before the market closed.

Because of the past success of TPE in which best estimates of people were aggregated without regard to time, we also explored averaging of estimates from people in the TPE condition cumulatively over days. This method showed only moderate accuracy, but we see that the representative dotted black line in Figure 1 is fairly similar to the other lines.

We had hypothesized that TPE would improve the prediction market's accuracy, but results suggest that after seeing the market history, people in the TPE condition behave similarly to people in the regular condition of the prediction market. Although we have already seen evidence that people are not anchoring on the state of the market when they make their edits, the typical influence of previous edits should not be seen as a bias because it does not decrease – and might even increase – the accuracy of forecasts.

ACKNOWLEDGMENTS
This research was supported by the Intelligence Advanced Research Projects Activity (IARPA) via Department of Interior National Business Center contract number D11PC20062. The U.S. Government is authorized to reproduce and distribute reprints for Governmental purposes notwithstanding any copyright annotation thereon. Disclaimer: The views and conclusions contained herein are those of the authors and should not be interpreted as necessarily representing the official
policies or endorsements, either expressed or implied, of IARPA, DoI/NBC, or the U.S. Government.

REFERENCES
H. J. Einhorn and R. M. Hogarth. 1978. Confidence in judgment: Persistence of the illusion of validity. *Psychological Review*, 86, 395-416.
J. Klayman, J. B. Soll, P. Juslin, and A. Winman. 2006. Subjective confidence and sampling of knowledge. Pp. 153–182 in Fiedler K, Juslin P (Eds). Information Sampling and Adaptive Cognition. New York, NY: Cambridge University Press.
A. Koriat, S. Lichtenstein, and B. Fischhoff. 1980. Reasons for confidence. *Journal of Experimental Psychology*, 6, 107-118.
S. E. Page. 2007. *The difference: How the power of diversity creates better groups, firms, schools, and societies*. Princeton, NJ : Princeton University Press.
P. W. Rhode and K. S. Strumpf. 2004. Historical presidential betting markets. *The Journal of Economic Perspectives*, 18, 127-141.
W. R. Sieck, E. C. Merkle, and T. Van Zandt. 2007. Option fixation: A cognitive contributor to overconfidence. *Organizational Behavior and Human Decision Processes*, 103, 68-83.
A. Speirs-Bridge, F. Fidler, M. McBride, L. Flander, G. Cumming, and M. Burgman. (2010). Reducing overconfidence in the interval judgments of experts. *Risk Analysis*, 30, 512-523.
J. B. Soll and J. Klayman. 2004. Overconfidence in interval estimates. *Journal of Experimental Psychology: Learning Memory and Cognition*, 30, 299–314.
J. Surowiecki. 2004. *The wisdom of crowds : Why the many are smarter than the few and how collective wisdom shapes business, economies, societies and nations*. New York, NY: Doubleday.
A. Tversky and D. Kahneman. 1974. Judgment under uncertainty: Heuristics and biases. *Science*, 185, 1124–1130.
T. D. Wilson, C. E. Houston, K. M. Etling, and N. Brekke. 1996. A new look at anchoring effects: Basic anchoring and its antecedents. *Journal of Experimental Psychology: General*, 125, 387-402.